\documentclass[a4paper,12pt]{article}

\usepackage{amsmath}
\usepackage{amssymb}
\usepackage{amsthm}
\usepackage{psfrag}
\usepackage{graphicx}
\usepackage{hyperref}
\usepackage{color}
\usepackage{bm}
\usepackage{float}

\newcommand{\be}{\begin{equation}}
\newcommand{\ee}{\end{equation}}
\newcommand{\LL}{\mathcal{L}}

\usepackage[left=3cm,right=2.5cm,top=2cm,bottom=3cm,bindingoffset=0cm]{geometry}

\usepackage{slashed}
\usepackage{amsmath}
\usepackage{amssymb}
\usepackage{amsthm}
\usepackage{psfrag}
\usepackage{graphicx}
\usepackage{slashed}
\usepackage[utf8]{inputenc}

\begin{document}

\title{
\begin{flushright}
{\small INR-TH-2016-050}
\end{flushright}
Electroweak vacuum stability in the Higgs--Dilaton theory}

\author{
A.\;Shkerin$^{a,b}$\thanks{{\bf e-mail}:
shkerin@inr.ru}
\\
$^a${\small{\em Institute of Physics, Ecole Polytechnique F\'ed\'erale de Lausanne (EPFL),
}}\\
{\small{\em  CH-1015, Lausanne, Switzerland}}\\
$^b${\small{\em
Institute for Nuclear Research of the Russian Academy
of Sciences,}}\\
{\small{\em 60th October Anniversary prospect 7a, 117312, Moscow,
Russia
}}
}
\date{}

\maketitle

\begin{abstract}
We study the stability of the Electroweak (EW) vacuum in a scale--invariant extension of the Standard Model and General Relativity, known as a Higgs--Dilaton theory. The safety of the EW vacuum against possible transition towards another vacuum is a necessary condition for the model to be phenomenologically acceptable. We find that, within a wide range of parameters of the theory, the decay rate is significantly suppressed compared to that of the Standard Model. We also discuss properties of a tunneling solution that are specific to the Higgs--Dilaton theory.
\end{abstract}

\section{Introduction}

An issue of stability of the electroweak (EW) vacuum is of considerable interest in recent years. In the Standard Model (SM), the tree--level Higgs potential has an absolute minimum corresponding to this vacuum. The quantum corrections modify the potential drastically through the Renormalization Group (RG) running of the Higgs quartic coupling $\lambda$ \cite{Bezrukov:2012sa,Degrassi:2012ry,Buttazzo:2013uya}, and a new minimum can develop at large scales, making the EW vacuum unstable. The shape of the Higgs potential at large scales is very sensitive to the SM parameters and, in particular, to the Higgs mass $m_H$ and the top quark mass $m_t$. At the moment, the largest uncertainty in the parameters of the potential is due to uncertainties in the top mass measurements \cite{Why-should-we-care}. The value of $m_t$ is extracted from the Monte--Carlo analysis of the decay products of the top quark, and it depends on the decay channels taken into account \cite{Spannagel:2016cqt,YauWong:2016idk}. Further uncertainties to the value of $m_t$ arise in theoretical analysis, where they are related to the difference between the Monte--Carlo and the pole masses of the quark. Note that these uncertainties leave the possibility for $\lambda$ to stay positive all the way up to the Planck scale, in which case no second minimum appears \cite{Why-should-we-care}. However, for the current best--fit values of the SM parameters, $\lambda$ crosses zero at the scale $\sim 10^{10}\,GeV$, and reaches its negative minimum at the scale $\sim 10^{17}\,GeV$ \cite{ATLAS:2014wva,Aad:2015zhl}. Hence, the possible instability of the EW vacuum must be properly taken care of, and a question of whether its life--time exceeds significantly the current age of the Universe deserves a special attention. 

The problem of stability of the EW vacuum is subject of numerous studies in the recent literature. These include the investigation of the vacuum decay in the SM Higgs potential and in flat space--time background \cite{Degrassi:2012ry,Buttazzo:2013uya}, the gravitational or thermal effects on the decay \cite{GravCorr-SM-1,GravCorr-SM-2,Rajantie:2016hkj,Salvio:2016mvj}, the vacuum stability at different cosmological epochs \cite{Cosmology-1,Cosmology-2,Cosmology-3,Cosmology-4} or in the presence of black holes \cite{BH-1,BH-2}. In this paper, we contribute to this research by analyzing the EW vacuum stability in one extension of the SM and General Relativity (GR) known as a Higgs--Dilaton theory \cite{HD-1,HD-2}.

The Higgs--Dilaton theory is an effective field theory whose properties allow it to account for many issues in particle physics and cosmology, which still lack of a complete explanation. For example, it makes a step towards a solution of the hierarchy problem by reformulating it in terms of dimensionless quantities. This is achieved by demanding the theory to be scale--invariant both at classical and quantum level. All scales are hence generated dynamically. Next, it is able to provide us with a plausible cosmological scenario, including inflation, dark matter and dark energy domination epochs of the Universe. Hence, the theory is phenomenologically acceptable in a wide range of scales. To confirm its viability, it remains to check whether the EW vacuum of the model is sufficiently safe compared to the SM case. We will be interested in the decay rate of the false vacuum, which is expressed as \cite{Coleman:1977py}
\be \label{DecayRate}
\Gamma=Ae^{-B}\,,~~~B=S_E(\text{bounce})-S_E(FV)\,.
\ee
Here $S_E(\text{bounce})$ is an euclidean action of the theory computed on a so--called bounce solution interpolating between the false and true vacua, $S_E(FV)$ is an euclidean action of the false vacuum (which is the EW vacuum in our case), and $A$ is a prefactor. In this paper, we will mainly focus on computing the exponential coefficient $B$. The value of $A$ can be simply estimated in the flat background \cite{Prefactor-flat-SM}, and we assume this approximation to hold in our case as well. We will find that in a wide range of parameters of the Higgs--Dilaton model, the EW vacuum decay probability is safely small. Moreover, the decay rate is suppressed significantly compared to that in the case of SM. We conclude that features of the Higgs--Dilaton model lead to additional stabilization of the false vacuum. 

We start in Section 2 with the brief review of the Higgs--Dilaton model, aiming to fix the notations and to introduce a particular set of field coordinates at which it is convenient to perform the analysis of classical solutions of the euclidean theory. In Section 3 we first provide an analytical investigation of the bounce solution and of an effective potential for it. We then find the bounce numerically and calculate the exponential factor $B$ for the wide range of parameters of the model. We discuss our findings and some further important properties of the bounce in Section 4. Finally, in Conclusion we collect the main results.

\section{Review of the Higgs--Dilaton model}

The Higgs--Dilaton model is a moderate extension of the SM and GR that contains no dimensional parameters at classical level. All scales are introduced as a result of a spontaneous breakdown of the scale invariance. The model contains the Higgs $\varphi$ and the dilaton $\chi$ fields coupled to gravity in a non--minimal way, and the rest of the SM content unchanged. The presence of the massless dilaton is necessary for the model to match the observational data \cite{HD-1}. The fields $\varphi$ and $\chi$ are allowed to interact in the way that preserves the scale invariance. The Lagrangian of the theory takes the form \footnote{We do not take into account possible boundary terms, since they do not affect the action on the bounce solution \cite{Gibbons:1976ue} (see also \cite{Masoumi:2016pqb}).}
\be\label{GeneralLagrangian}
\dfrac{\LL}{\sqrt{-g}}=\dfrac{1}{2}(2\xi_h\varphi^\dag\varphi+\xi_\chi\chi^2)R-\dfrac{1}{2}g^{\mu\nu}\partial_\mu\chi\partial_\nu\chi - V(\varphi,\chi)+\LL_{SM,\lambda\rightarrow 0}\,,
\ee
where the last term is the SM Lagrangian without the Higgs potential. We will refer to the Lagrangian (\ref{GeneralLagrangian}) as the one written in the Jordan ($J-$)frame. The potential $V(\varphi,\chi)$ is taken in the most general form, compatible with the scale invariance,
\be\label{GeneralPotential}
V(\varphi,\chi)=\lambda\left(\varphi^\dag\varphi-\dfrac{\alpha}{2\lambda}\chi^2\right)^2+\beta\chi^4\,,
\ee
with $\alpha$ and $\beta$ some constants.

The requirement for the model to be phenomenologically viable constrains its parameters significantly. First, note that the non--minimal couplings $\xi_h$ and $\xi_\chi$ must be positive. This ensures the semi--positive definiteness of kinetic terms for the scalar fields $\varphi$ and $\chi$. Requiring further for the theory to yield a successful inflationary scenario restricts the couplings to the values $\xi_h\sim 10^3-10^6$, and $\xi_\chi\lesssim 10^{-2}$ \cite{HD-1}. In this paper, we relax this condition by allowing the couplings to be simply 
\be \label{ConditionsOnXi}
\xi_\chi\ll 1\ll\xi_h\,.
\ee
Next, the Higgs--Dilaton model reproduces the observed hierarchy between the Planck, electroweak and cosmological scales provided that 
\be \label{ConditionsOnAB}
\beta\lll\alpha\lll 1\,,~~~\alpha\sim\beta^{1/4}\ll\xi_\chi\,.
\ee
For our purposes, one can safely neglect the contributions from the corresponding terms of the potential and set $\alpha=\beta=0$. Indeed, such approximation is clearly applicable as long as $\lambda \vert\varphi\vert^2 \gg\alpha M_P^2$. However, as will be shown later, the contribution to the decay exponent from the region of $\vert\varphi\vert$ where this condition violates is itself negligible if the inequalities (\ref{ConditionsOnXi}) and (\ref{ConditionsOnAB}) hold. Hence, the approximation is justified for all values of $\vert\varphi\vert$. Choosing the unitary gauge for the Higgs field, $\varphi^T=(0,h/\sqrt{2})$, we rewrite the potential (\ref{GeneralPotential}) as follows,
\be\label{JordanFramePotential}
V(h)=\dfrac{\lambda}{4}h^4\,.
\ee
Finally, the euclidean form of the Lagrangian (\ref{GeneralLagrangian}) is written as
\be 
\dfrac{\LL_E}{\sqrt{g}}=-\dfrac{1}{2}(\xi_\chi\chi^2+\xi_hh^2)R+\dfrac{1}{2}(\partial h)^2 +\dfrac{1}{2}(\partial\chi)^2+V(h)\,,
\ee
where we omitted all terms in $\mathcal{L}_{SM,\lambda\rightarrow 0}$, except the kinetic term for the Higgs field, since they are  not relevant for our analysis. Whenever the non--minimal couplings are non--zero, one can perform the following metric redefinition,
\be 
\tilde{g}_{\mu\nu}=\Omega^2 g_{\mu\nu}\,,
\ee
to rewrite the theory in the form in which the non--minimal couplings are absent. We will refer to this form as the Einstein ($E-$)frame. To achieve this, the conformal factor $\Omega^2$ has to be chosen as
\be 
\Omega^2=M_P^{-2}(\xi_hh^2+\xi_\chi\chi^2)\,.
\ee
Making use of the standard relations between the $J-$ and $E-$frames \cite{Fujii:2003pa},
\be 
\sqrt{g}=\Omega^{-4}\sqrt{\tilde{g}}\,,~~~ R=\Omega^2(\tilde{R}+6\tilde{\square}\log\Omega-6\tilde{g}^{\mu\nu}\partial_\mu\log\Omega\partial_\nu\log\Omega)\,,
\ee
rewritten in the euclidean signature, we arrive at the following Lagrangian
\be\label{EframeLagrangian}
\dfrac{\LL_E}{\sqrt{\tilde{g}}}=-\dfrac{M_P^2}{2}\tilde{R}+\dfrac{1}{2}\tilde{K}(h,\chi)+\tilde{V}(h,\chi)\,.
\ee
The kinetic term $\tilde{K}$ has a non--canonical form,
\be 
\tilde{K}(h,\chi)=\gamma_{ij}\tilde{g}^{\mu\nu}\partial_\mu\phi^i\partial_\nu\phi^j\,,
\ee
where we introduced the notation $(\phi^1,\phi^2)\equiv (h,\chi)$. The quantity $\gamma_{ij}$ can be interpreted as a metric in the two--dimensional field space spanned by $h$ and $\chi$, in the $E-$frame. It is given by
\be \label{FieldSpaceMetric}
\gamma_{ij}\equiv\dfrac{1}{\Omega^2}\left(\delta_{ij}+\dfrac{3}{2}M_P^2\dfrac{\partial_i\Omega^2\partial_j\Omega^2}{\Omega^2}\right)\,.
\ee
Finally, the transformed potential is written as
\be 
\tilde{V}(h,\chi)=\dfrac{V(h)}{\Omega^4}\,.
\ee
We now look for further redefinition of the fields of the theory, aiming to recast the field space metric (\ref{FieldSpaceMetric}) into a diagonal form. To this end, we exploit the scale invariance of the model. Consider the infinitesimal scale transformation of the fields in the $E-$frame,
\be 
\tilde{g}_{\mu\nu}\rightarrow\tilde{g}_{\mu\nu}\,,~~~\phi^i\rightarrow\phi^i+\sigma\Delta\phi^i\,,
\ee
where $\sigma$ is a small constant. The current corresponding to this transformation reads as follows,
\be \label{CurrentOld}
\tilde{J}^\mu=\dfrac{1}{\sqrt{\tilde{g}}}\dfrac{\partial\LL_E}{\partial\partial_\mu\phi^i}\Delta\phi^i=\tilde{g}^{\mu\nu}\dfrac{M_P^2}{2(\xi_\chi\chi^2+\xi_hh^2)}\partial_\nu((1+6\xi_\chi)\chi^2+(1+6\xi_h)h^2)\,.
\ee
Following \cite{HD-1}, we introduce a new set of variables $(\phi'^1,\phi'^2)\equiv (\rho,\theta)$ that transform under the scale transformations as

\be\label{ScaleTransformLaw}
\rho\rightarrow\rho + \sigma M_P\,,~~~\theta\rightarrow\theta\,.
\ee

Due to the scale invariance of the Lagrangian (\ref{EframeLagrangian}), the field $\rho$ can only enter the Lagrangian through its derivatives. Requiring the metric $\gamma'_{ij}$ corresponding to the fields $(\rho,\theta)$ to be diagonal, we have
\be \label{CurrentNew}
\tilde{J}^\mu=M_P\tilde{g}^{\mu\nu}\gamma'_{\rho\rho}\partial_\nu\rho\,.
\ee
Comparing the currents (\ref{CurrentOld}) and (\ref{CurrentNew}), we deduce the following expression for $\rho$,
\be \label{Rho}
\rho=\dfrac{M_P}{2}\log\left(\dfrac{(1+6\xi_\chi)\chi^2+(1+6\xi_h)h^2}{M_P^2}\right)\,.
\ee
We observe that $\rho$ can be viewed as a radial coordinate in the field space spanned by the vectors $\sqrt{1+6\xi_\chi}\chi$ and $\sqrt{1+6\xi_h}h$. We can choose $\theta$ to be an angular coordinate in this space, that is
\be \label{Theta}
\theta=\arctan\left(\sqrt{\dfrac{1+6\xi_h}{1+6\xi_\chi}}\dfrac{h}{\chi}\right)\,.
\ee
By construction, $\theta$ does not transform under the scale transformations, in agreement with (\ref{ScaleTransformLaw}). In terms of $\theta$ and $\rho$, the Lagrangian (\ref{EframeLagrangian}) is written as
\be\label{NewLagrangian}
\dfrac{\LL_E}{\sqrt{g}}=-\dfrac{M_P^2}{2}\tilde{R}+\dfrac{a(\theta)}{2}(\partial\rho)^2+\dfrac{b(\theta)}{2}(\partial\theta)^2+\tilde{V}(\theta)\,,
\ee
with the potential
\be\label{PotentialForTheta}
\tilde{V}(\theta)=\dfrac{\lambda}{4\xi_h^2}M_P^4\left(\dfrac{1}{1+\varsigma\cot^2\theta}\right)^2\,,
\ee
where
\be 
a(\theta)=\dfrac{1+6\xi_h}{\xi_h}\dfrac{1}{\sin^2\theta+\varsigma\cos^2\theta}\,,~~~b(\theta)=\dfrac{M_P^2\varsigma}{\xi_\chi}\dfrac{\tan^2\theta+\xi_\chi/\xi_h}{\cos^2\theta(\tan^2\theta+\varsigma)^2}\,,
\ee
and
\be \label{varsigma}
\varsigma=\dfrac{(1+6\xi_h)\xi_\chi}{(1+6\xi_\chi)\xi_h}\,.
\ee
We see that the fields $\rho$ and $\theta$ are almost decoupled, and, what is more important, the potential $\tilde{V}$ depends on $\theta$ only. This simplifies significantly the study of the classical solutions in the theory.

\section{Bounce in the Higgs--Dilaton model}

\subsection{Equations of motion and boundary conditions}

A bounce is a regular solution interpolating between the regions of true and false vacua. Since we study the vacuum decay in a homogeneous and isotropic environment, we can assume the bounce to be $O(4)-$symmetric.\footnote{Although it was proven that the solution of maximal symmetry dominates the transition amplitude in flat space background \cite{O4-flat,Blum:2016ipp}, no such proof is known in the case when gravity dynamics is included.} Hence, the following ansatz for the metric can be chosen, 
\be \label{MetricAnsatz}
d\tilde{s}^2=g^2(r)dr^2+r^2d\Omega_3^2\,,
\ee
where $r$ is a radial coordinate, and $d\Omega_3^2$ denotes the angular part of the metric. In what follows, we neglect the space--time curvature arising due to the current value of the cosmological constant $\Lambda_0$, by assuming the false vacuum state geometry to be flat. As will be shown later, this is a reasonable approximation as long as $m_H\gg\Lambda_0^{1/4}$. In this case, the function $g$ is required to approach the flat space limit at infinity and the euclidean AdS limit at the origin. The scalar fields $\rho$ and $\theta$ are required to have a good behavior at infinity, in order to ensure the finiteness of the action, and to be regular at the origin. 

Applying the ansatz (\ref{MetricAnsatz}) to the equations of motion, following from the Lagrangian (\ref{NewLagrangian}), one finds,
\be \label{SolutionForRho}
\rho'=C\cdot\dfrac{g}{a(\theta)r^3}\,,
\ee
with $C$ some constant. It is easy to see that the requirement for $\theta$ to approach a finite true vacuum value $\theta_0$ at the origin spoils whenever $C\neq 0$. Hence, the tunneling solution must obey $\rho=\rho_0=\text{const}$, and the value of $\rho_0$ is fixed by the false vacuum state, $h_{FV}=0$, $\chi_{FV}=M_P/\sqrt{\xi_\chi}$ \cite{HD-1}.\footnote{We neglect corrections due to the non--zero $\alpha$ and $\beta$.} From Eq.~(\ref{Rho}) we have
\be \label{BounceRho}
\rho_0=\dfrac{M_P}{2}\log\left(\dfrac{1+6\xi_\chi}{\xi_\chi}\right)\,.
\ee 
Under the conditions (\ref{MetricAnsatz}) and (\ref{BounceRho}), the equations of motion become
\be\label{EqOnTheta}
2rg^3\tilde{V}'(\theta)+2rb(\theta)g'\theta'-g(rb'(\theta)\theta'^2+2b(\theta)(3\theta'+r\theta''))=0\,,
\ee
\be\label{EqOnG}
g^2=\dfrac{6M_P^2-r^2b(\theta)\theta'^2}{2(3M_P^2-r^2\tilde{V}(\theta))}\,.
\ee
The system (\ref{EqOnTheta}), (\ref{EqOnG}) is to be solved numerically. However, before plunging into numerics, it is useful to understand the qualitative behavior and asymptotic properties of the bounce analytically. This analysis is performed below.

\subsection{Running couplings}

In the Quantum theory framework, the potential (\ref{PotentialForTheta}) gets modified due to the RG running of the coupling constants $\xi_h$, $\xi_\chi$ and $\lambda$. Quantum theory predictions are extracted from the classical action of the theory and from the set of subtraction rules used to renormalize it. The Higgs--Dilaton model, whose classical Lagrangian is given by (\ref{NewLagrangian}), is not renormalizable, hence the physical results will depend on the way we regularize it. We choose to regularize the quantum theory in such a way that all the symmetries of the classical action, including scale invariance, remain intact. An example of such regulatization procedure was constructed in \cite{SI-renorm} (see also \cite{SI-renorm-2}), and applied to the Higgs--Dilaton model in \cite{HD-2}. It is based on dimensional regularization, in which the 't Hooft--Veltman normalization point $\mu$ is replaced by some function of the fields $h$, $\chi$, in the $J-$frame. The different choices of the function produce different physical results. We will consider two most natural possibilities. One of them corresponds to identification of $\mu$ with the gravitational cut--off in the $J-$frame,
\be 
\mu_I\sim \xi_\chi \chi^2+\xi_hh^2\,.
\ee
The second possibility is to choose the scale--invariant direction along the dilaton field, i.e.,
\be 
\mu_{II}\sim \xi_\chi\chi^2\,.
\ee
To test the ability of the Higgs--Dilaton model to describe correctly the inflationary physics, the careful analysis of the quantum corrections to the potential (\ref{JordanFramePotential}) during inflation is needed. Such analysis was performed in \cite{HD-2}. It was shown that at one--loop level the leading contribution to the potential is given by
\be \label{LogContr}
\Delta V=-\dfrac{3m_t^4}{16\pi^2}\left(\log\dfrac{m_t^2}{\mu^2}-\dfrac{3}{2}\right)\,,
\ee
where $m_t^2=y_t^2h^2/2$ stands for the effective top mass in the $J-$frame. One can now replace $\mu^2=\frac{\hat{\mu}^2}{M_P^2}F(h,\chi)$ and treat $\hat{\mu}^2$ as a usual momentum scale on which nothing depends in the final result, since the change of it would be compensated by the running of $\lambda(\hat{\mu})$ and $\xi_{h,\chi}(\hat{\mu})$. It is convenient to fix the value of $\hat{\mu}$ such that the logarithmic contribution (\ref{LogContr}) is minimized for each value of $h$, $\hat{\mu}^2\simeq\frac{y_t^2}{2}\frac{h^2}{F(h,\chi)/M_P^2}$. Depending on the choice of the normalization point $\mu_{I,II}$, this gives,
\be 
\hat{\mu}^2_I(h,\chi)=\dfrac{y_t^2}{2}\dfrac{M_P^2h^2}{\xi_hh^2+\xi_\chi\chi^2}\,,~~~\hat{\mu}^2_{II}(h,\chi)=\dfrac{y_t^2}{2}\dfrac{M_P^2h^2}{\xi_\chi\chi^2}\,.
\ee
Finally, we rewrite these expressions in terms of the variables $\rho$ and $\theta$ to obtain
\be\label{NormalizationPoints}
\hat{\mu}_I^2(\theta)=\dfrac{y_t^2}{2}\dfrac{M_P^2}{\xi_h}\dfrac{\sin^2\theta}{1-(1-\varsigma)\cos^2\theta}\,,~~~\hat{\mu}_{II}^2(\theta)=\dfrac{y_t^2}{2}\dfrac{M_P^2}{\xi_h\varsigma}\cot^2\theta\,,
\ee
with $\varsigma$ given in (\ref{varsigma}). In accordance with the chosen regularization scheme, the momentum scale depends only on the scale invariant quantity $\theta$. The RG enhanced potential for the field $\theta$ is given by (\ref{PotentialForTheta}) with $\lambda$ replaced by the running coupling $\lambda(\hat{\mu}_{I,II}(\theta))$.\footnote{It what follows, we neglect the running of the couplings $\xi_h$, $\xi_\chi$.}

\subsection{Effective potential}

\begin{figure}[h]
\begin{center}
\begin{minipage}[h]{0.4\linewidth}
\center{\includegraphics[width=1\linewidth]{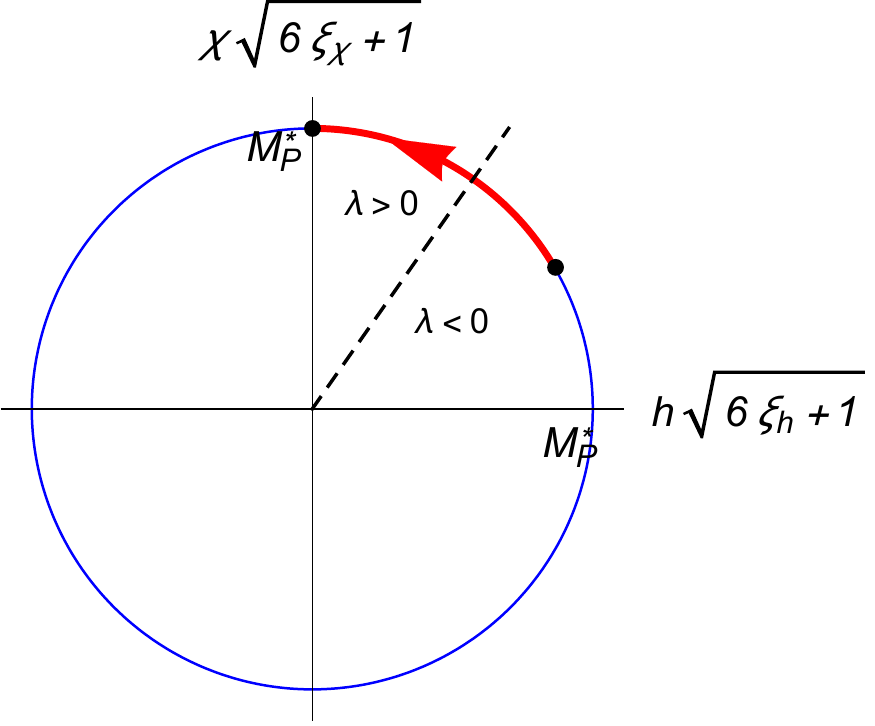}}
\end{minipage}
\caption{The bounce configuration in terms of the Higgs ($h$) and dilaton ($\chi$) fields. An arrow points the direction in which $r$ grows.}
\label{Plot:Ellipse}
\end{center}
\end{figure}

To get an insight into qualitative properties of the bounce, it is useful to rewrite the solution $\rho=\rho_0$, $\theta=\theta_b(r)$ in terms of the original variables $h$ and $\chi$. From the formula (\ref{Rho}) we obtain the relation between $h_b$ and $\chi_b$,
\be \label{BounceEllipse}
(1+6\xi_\chi)\chi_b^2+(1+6\xi_h)h_b^2=M_P^{*2}\,,~~~M_P^*=M_P\sqrt{\dfrac{1+6\xi_\chi}{\xi_\chi}}\,.
\ee
One observes that the bounce trajectory draws a circle in the field space spanned by the vectors $\sqrt{1+6\xi_\chi}\chi$ and $\sqrt{1+6\xi_h}h$, as shown in Fig.(\ref{Plot:Ellipse}). The relation (\ref{BounceEllipse}) allows us to study the bounce using a single variable which we choose to be $h_b$. Using Eqs.~(\ref{BounceEllipse}) and (\ref{Theta}), one finds the relation between $h_b$ and $\theta_b$,
\be \label{h(theta)}
h_b=\dfrac{M_P^*}{\sqrt{1+6\xi_h}}\sin\theta_b\,.
\ee
By the definition (\ref{Theta}), $\theta_b$ is confined in the interval $0\leqslant\theta_b\leqslant\frac{\pi}{2}$. This condition, seeming obscuring in the $(\rho,\theta)-$variables, becomes clear if we write it in terms of $h_b$,
\be \label{BounceBound}
0\leqslant h_b\leqslant\dfrac{M_P^*}{\sqrt{1+6\xi_h}}\,,
\ee
where it is seen to be the consequence of Eq.~(\ref{BounceEllipse}). The inequality (\ref{BounceBound}) imposes a non--trivial condition on the magnitude $h_0$ of the bounce. We will say more about this below.

Using Eqs.~(\ref{PotentialForTheta}) and (\ref{BounceEllipse}), one obtains the effective potential for the bounce,
\be\label{EffectivePotential}
V_{\textit{eff}}=\dfrac{\lambda(\hat{\mu}(h_b))}{4}\left(\dfrac{1}{M_P^2}\dfrac{\xi_h-\xi_\chi}{1+6\xi_\chi}+\dfrac{1}{h_b^2}\right)^{-2}\,.
\ee
One minimum of this potential is achieved at $h_b=0$, in accordance with the false vacuum solution $\theta_{FV}=0$ of the equation of motion (\ref{EqOnTheta}).\footnote{We neglect corrections due to the non--zero vev of the Higgs field.} Another, deeper minimum develops whenever $\lambda(\hat{\mu}_{I,II}(h_b))$ crosses zero at some scale $h_*$. Note also that, as long as the condition (\ref{ConditionsOnXi}) is fulfilled, the potential (\ref{EffectivePotential}) possesses no singular points. 

Now we would like to investigate how variations of the different couplings, that are present in the potential (\ref{EffectivePotential}), affect the decay rate (\ref{DecayRate}). Presumably, the strongest effect on the bounce is caused by the variation of the Higgs quartic coupling $\lambda(\hat{\mu})$. For this reason, below we choose different values of the top mass $m_t$ --- those, lying in the $2\sigma$ experimental uncertainty region according to the combined fit provided in \cite{ATLAS:2014wva}. However, the effect of varying $m_t$ is well--known in the case of SM (see the Introduction), and we do not expect it to change much in the Higgs--Dilaton theory.

Next, we turn to the non--minimal couplings $\xi_h$ and $\xi_\chi$. We ask what the signs of variations $\frac{\delta B}{\delta\xi_h}$ and $\frac{\delta B}{\delta\xi_\chi}$ are in the cases when the normalization prescriptions (\ref{NormalizationPoints}) are implemented.

\textbf{Prescription $I$.} As will be seen later from numerics, under the conditions (\ref{ConditionsOnXi}) the magnitude of the bounce satisfies 
\be \label{BoundBounceMagnitude}
\dfrac{h_0^2\xi_h}{M_P^2}\ll 1\,.
\ee
Then, using Eq.~(\ref{h(theta)}), one expresses the normalization point $\hat{\mu}_I$ through $h_b$ as follows,
\be \label{FirstNorm}
\hat{\mu}^2_I(h_b)=\dfrac{y_t^2h_b^2}{2}\left(1-\dfrac{\xi_h-\xi_\chi}{1+6\xi_\chi}\dfrac{h_b^2}{M_P^2}+O\left(\dfrac{h_b^4\xi_h^2}{M_P^4}\right)\right)\,.
\ee
If $h_*\ll h_0$, one expects the dominant contribution to the bounce action coming from the region of $r$ at which the bounce solution is determined mainly by the behavior of the effective potential $V_{\textit{eff}}$ at large $h_b$. Hence, the variation of $B$ is determined by the variation of the asymptotics of $V_{\textit{eff}}$. From Eqs.~(\ref{EffectivePotential}) and (\ref{FirstNorm}) we have,
\be \label{ChangeOfV}
\dfrac{\delta\vert V_{\textit{eff}}\vert}{\delta\xi_h}<0\,,~~~\dfrac{\delta\vert V_{\textit{eff}}\vert}{\delta\xi_\chi}>0\,,
\ee
from which it follows that
\be \label{ChangeOfBI}
\dfrac{\delta B}{\delta\xi_h}>0\,,~~~\dfrac{\delta B}{\delta\xi_\chi}<0\,.
\ee

\begin{figure}[h]
\begin{center}
\begin{minipage}[h]{0.4\linewidth}
\center{\includegraphics[width=1\linewidth]{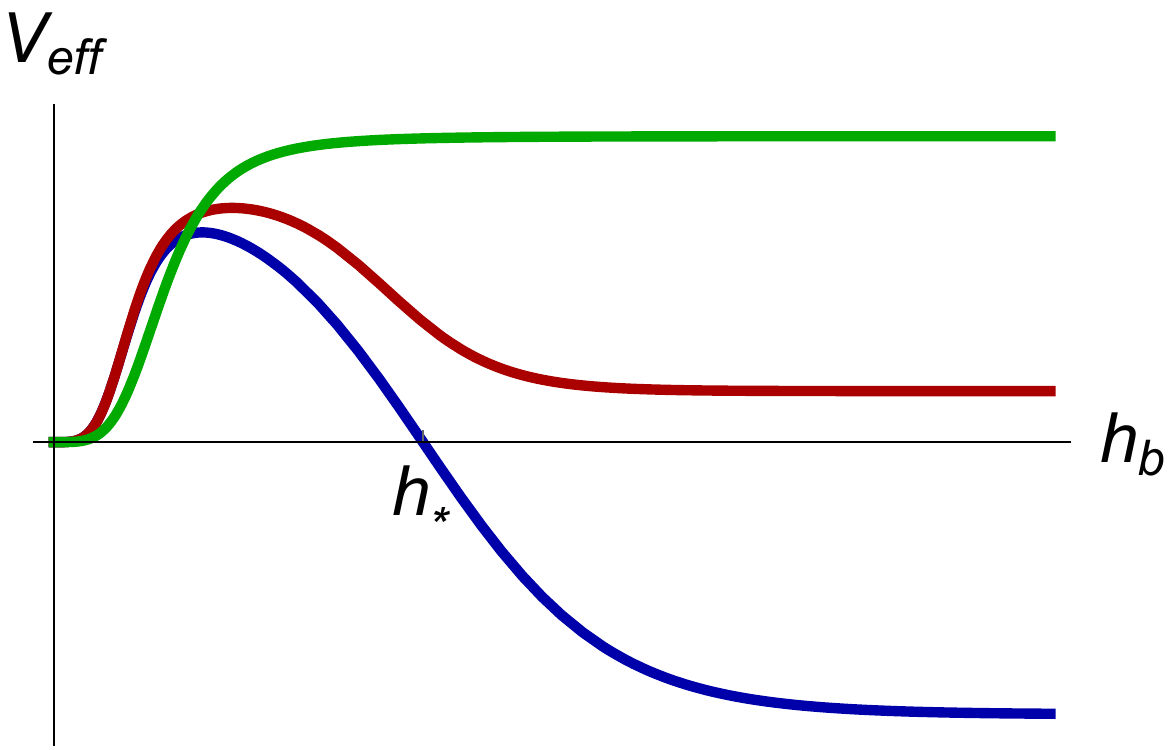}}
\end{minipage}
\caption{Possible forms of the effective potential for the field $h_b$.}
\end{center}
\label{Plot:EffPot}
\end{figure}

\textbf{Prescription $II$.} Under the condition (\ref{BoundBounceMagnitude}), the dependence of the normalization scale $\hat{\mu}_{II}$ on the variable $h_b$ can be written as
\be \label{SecondNorm}
\hat{\mu}^2_{II}(h_b)= \dfrac{y_t^2h_b^2}{2}\left(1+\dfrac{\xi_\chi(1+6\xi_h)}{1+6\xi_\chi}\dfrac{h_b^2}{M_P^2}+O\left(\dfrac{h_b^4\xi_h^2\xi_\chi^2}{M_P^4}\right)\right)\,.
\ee
Using Eqs.~(\ref{EffectivePotential}) and (\ref{SecondNorm}), we arrive again at the result (\ref{ChangeOfV}), from which the inequalities (\ref{ChangeOfBI}) follow. Thus, we expect that for both normalization prescriptions the exponential coefficient $B$ grows as $\xi_h$ increases or $\xi_\chi$ decreases, making the tunneling less probable. These expectations are confirmed by numerical results.

\subsection{Decay rate}

By making use of the Einstein equations, the euclidean action on the bounce can be brought to the form
\be \label{BounceAction}
S_E(\text{bounce})\simeq-2\pi^2\int_0^{m_H^{-1}} dr~r^3g_b(r)\tilde{V}(\theta_b(r))\,,
\ee
where $g_b$ and $\theta_b$ are the bounce solution of Eqs.~(\ref{EqOnTheta}), (\ref{EqOnG}). The integral is truncated from above by the non--zero Higgs mass. Indeed, as long as $r\ll m_H^{-1}$, the Higgs field is effectively massless, and the bounce exhibits a power--like asymptotics, $\theta_b\sim r^{-2}$, that contributes to the integral (\ref{BounceAction}). At $r\gtrsim m_H^{-1}$, the bounce becomes decaying exponentially fast, and the contribution to the action from that region of $r$ is negligible. This allows us to justify the approximation that we made for the potential $V(h,\chi)$. Namely, as long as $m_H^2\gg\alpha M_P^2/\lambda$, the corrections to the bounce coming from the non--zero $\alpha$ and $\beta$ can be neglected. It was shown in \cite{HD-1} that under the conditions (\ref{ConditionsOnXi}) and (\ref{ConditionsOnAB}), the Higgs mass is given by
\be 
m_H^2\sim\dfrac{\alpha M_P^2}{\lambda\xi_\chi}\,,
\ee
hence the required inequality is fulfilled. We can also justify the flat space approximation for the false vacuum state that we made when discussing the boundary conditions for the bounce solution. Indeed, as long as $m_H\gg\Lambda_0^{1/4}$, the integral (\ref{BounceAction}) is insensitive to the non--trivial space geometry, and the flat space asymptotics can be used.

\begin{figure}[H]
\begin{minipage}[h]{0.44\linewidth}
\center{(pr.$I$, $m_{t}^-$)}
\center{\includegraphics[width=1\linewidth]{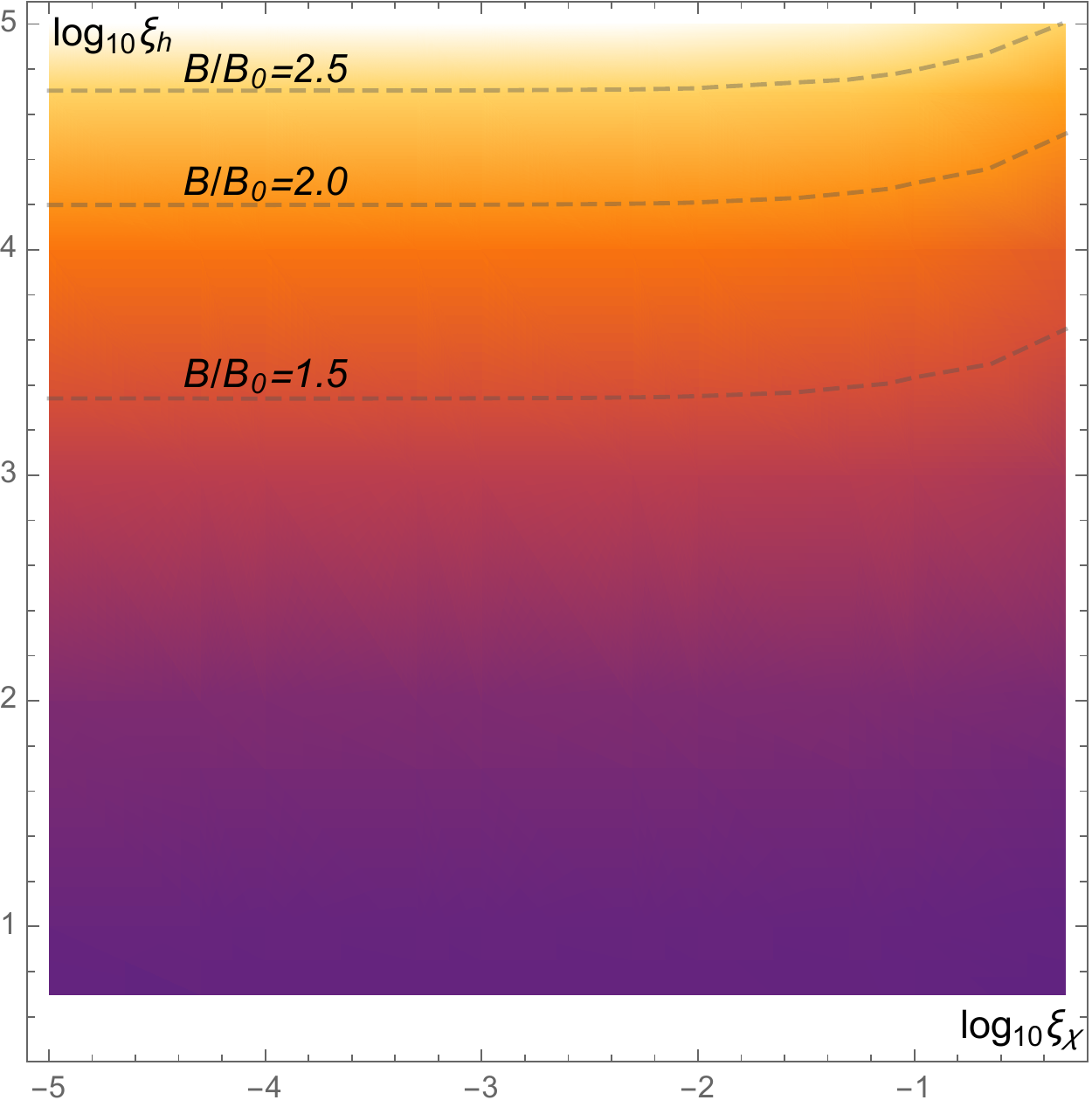}}
\end{minipage}
\hfill
\begin{minipage}[h]{0.44\linewidth}
\center{(pr.$II$, $m_{t}^-$)}
\center{\includegraphics[width=1\linewidth]{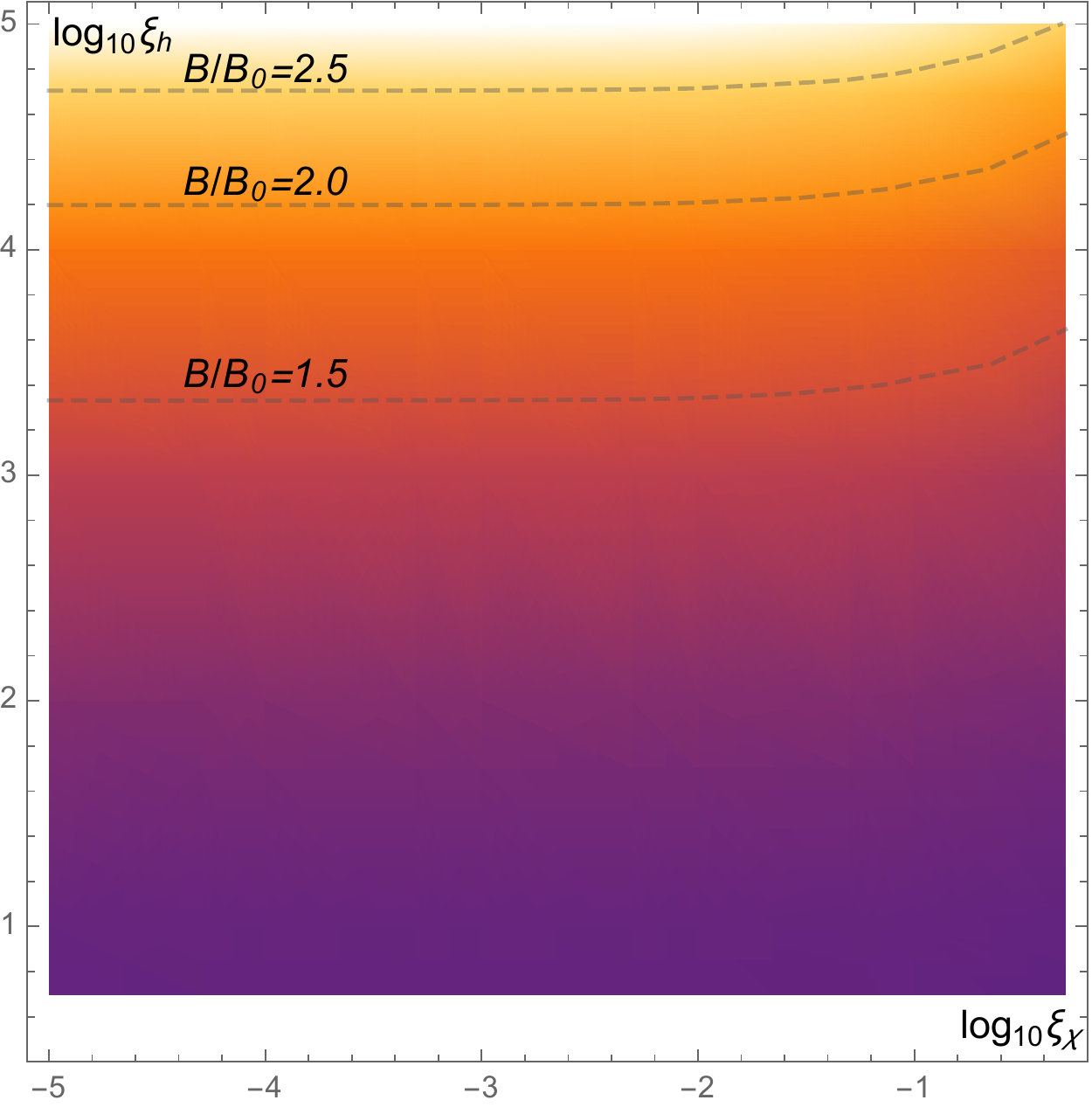}}
\end{minipage}

\begin{minipage}[h]{0.44\linewidth}
\center{(pr.$I$, $m_{t}$)}
\center{\includegraphics[width=1\linewidth]{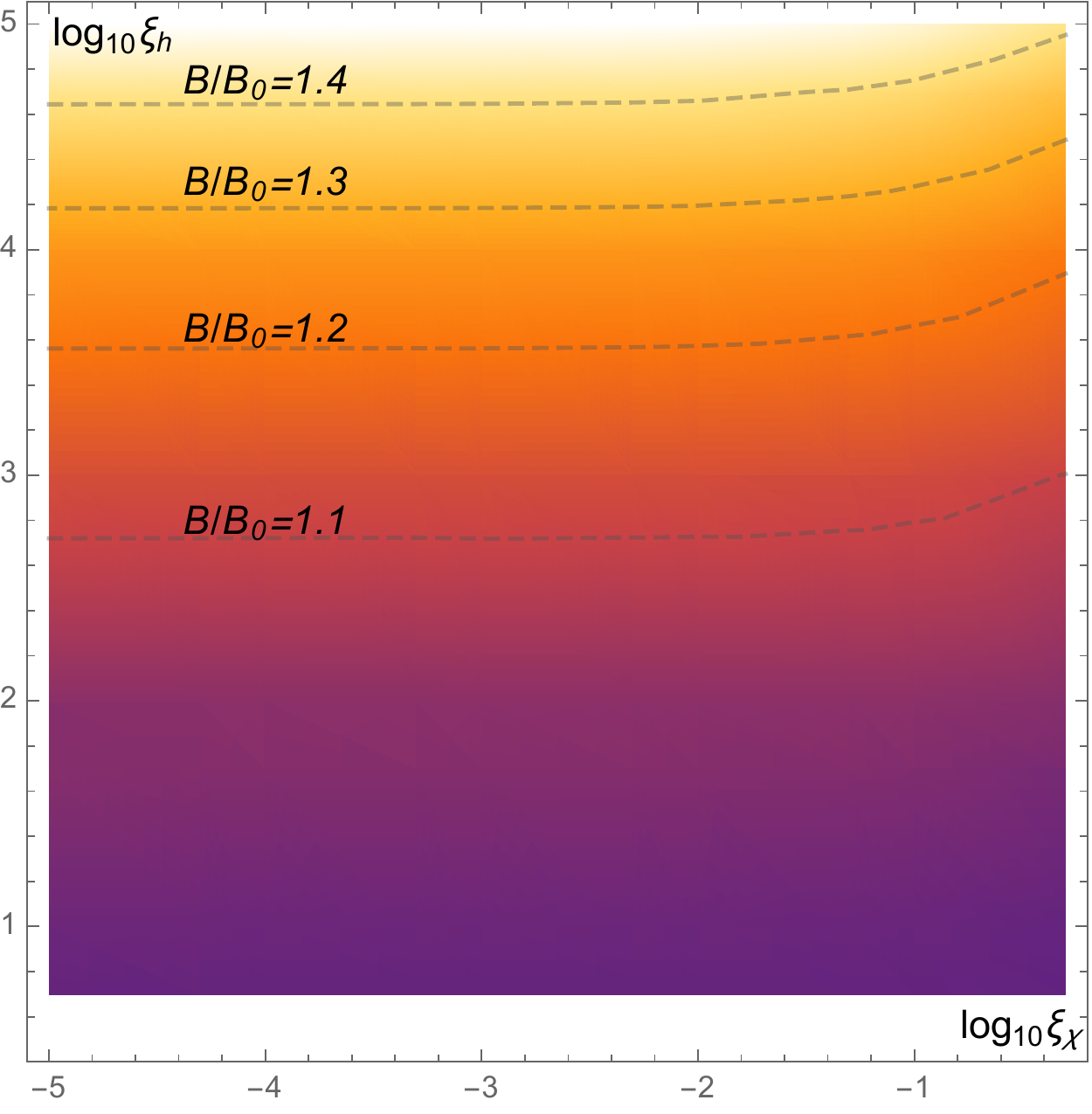}} 
\end{minipage}
\hfill
\begin{minipage}[h]{0.44\linewidth}
\center{(pr.$II$, $m_{t}$)}
\center{\includegraphics[width=1\linewidth]{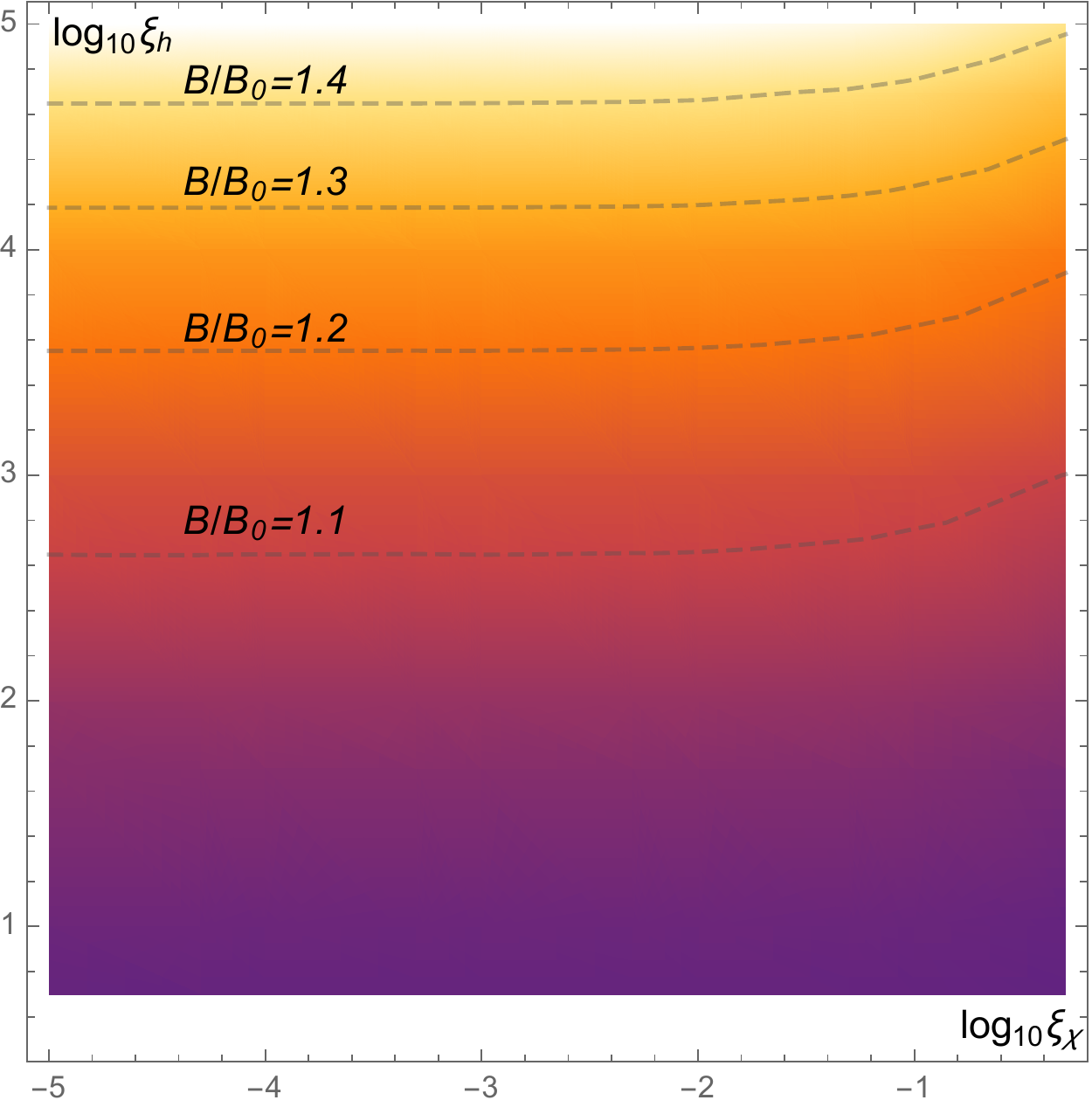}}
\end{minipage}

\begin{minipage}[h]{0.44\linewidth}
\center{(pr.$I$, $m_{t}^+$)}
\center{\includegraphics[width=1\linewidth]{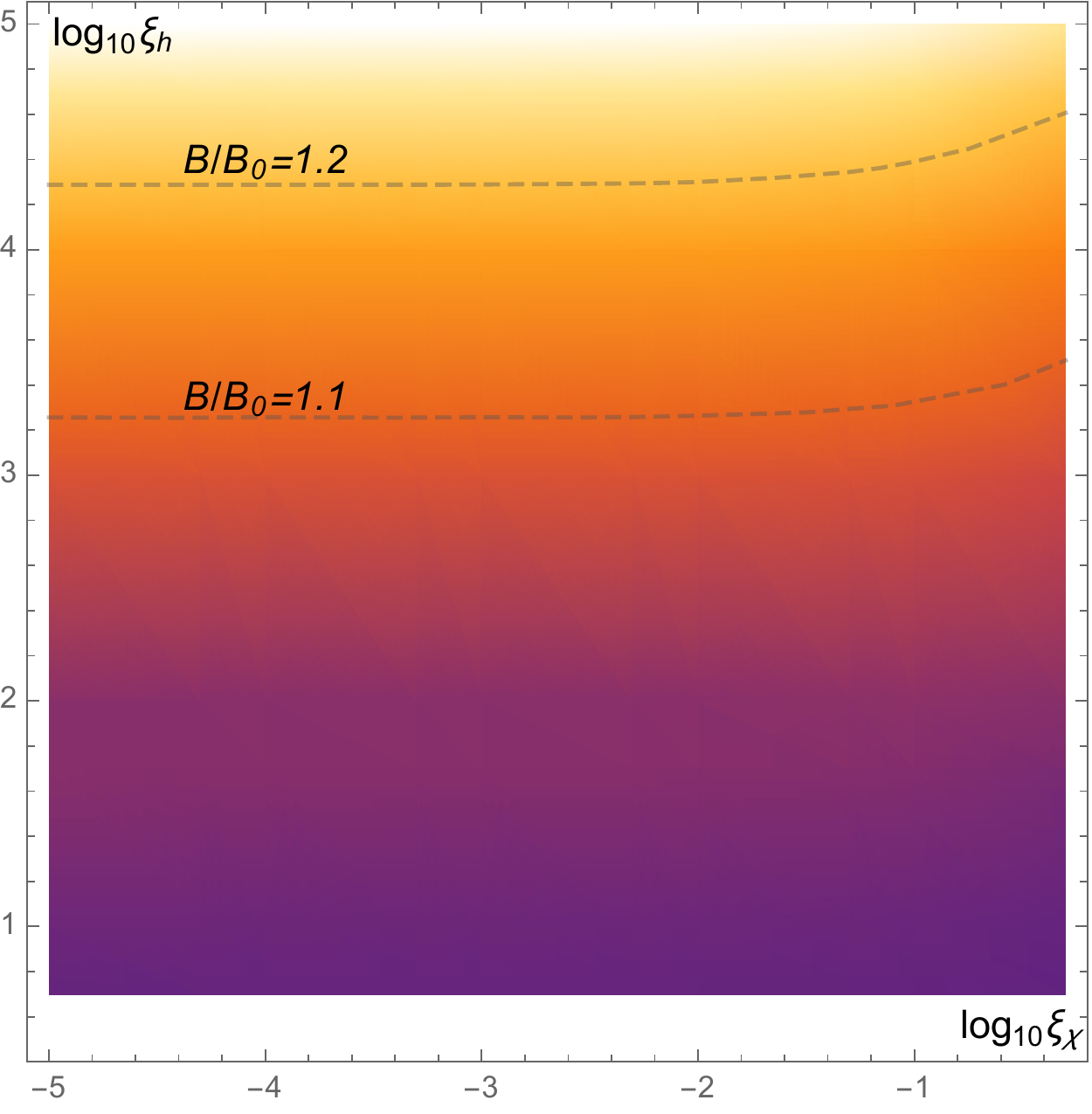}}
\end{minipage}
\hfill
\begin{minipage}[h]{0.44\linewidth}
\center{(pr.$II$, $m_{t}^+$)}
\center{\includegraphics[width=1\linewidth]{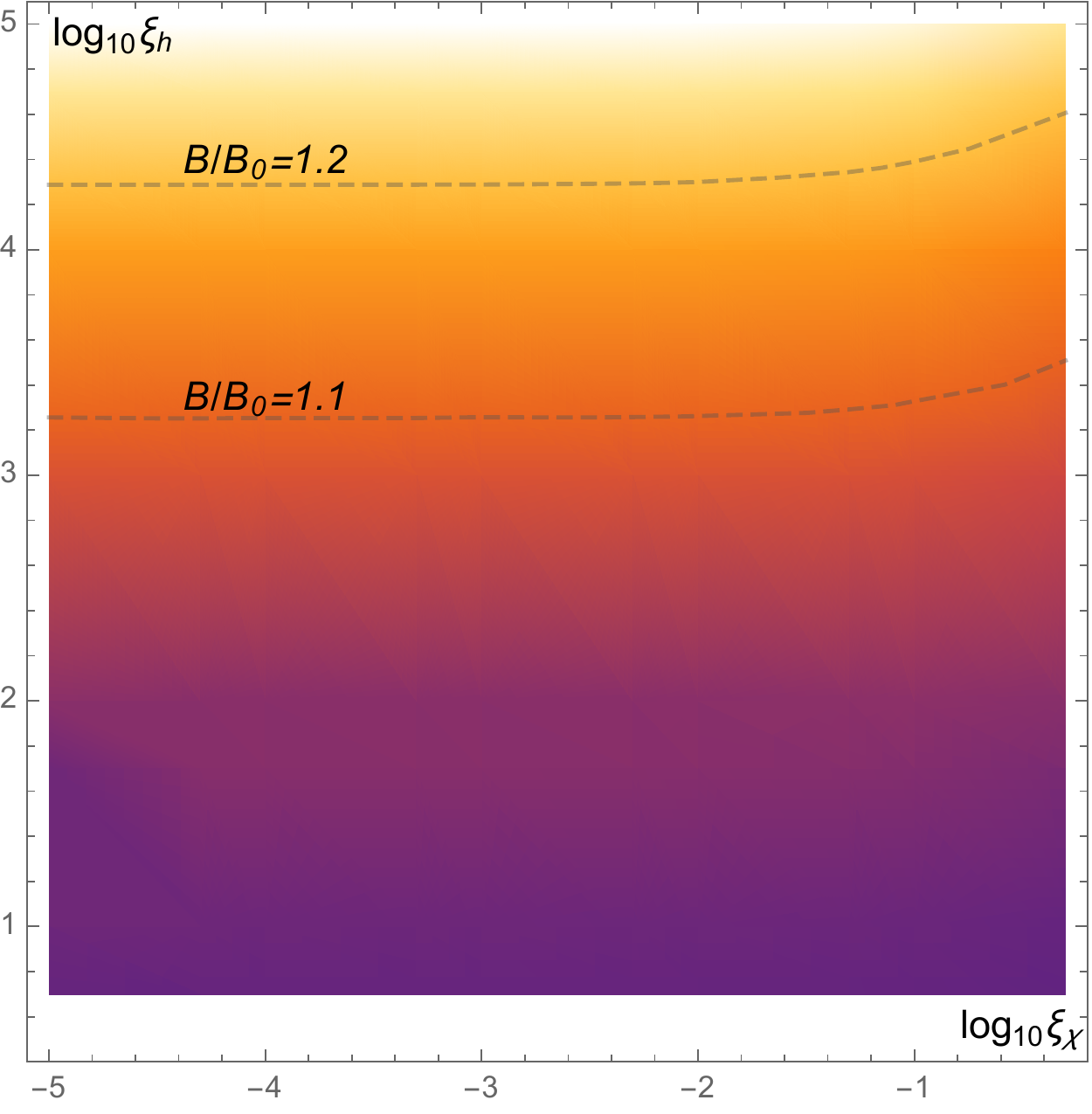}}
\end{minipage}
\caption{(see the text below) The ratio $B/B_0$ for the two choices of the normalization point (pr.$I$,$II$). We take the Higgs mass $m_H=125.09\,GeV$ \cite{Aad:2015zhl}, and the top masses $m_t=173.34\,GeV$ and $m_t^{\pm}=173.34\pm 1.52\,GeV$, corresponding to the $2\sigma$ experimental uncertainty region \cite{ATLAS:2014wva}.}
\label{Plot:Bs}
\end{figure}

Now we turn to the calculation of the decay rate (\ref{DecayRate}), focusing on the exponential coefficient $B$. In our case, $S_E(FV)=0$, and $B$ is given by the r.h.s. of Eq.~(\ref{BounceAction}). We are interested in the ratio $B/B_0$, where $B_0$ is the SM bounce action in flat space and for the same values of the SM parameters. To compute the RG running of the Higgs quartic coupling $\lambda$, we use the code based on \cite{RG-code-based,Bezrukov:2012sa}. We take the Higgs mass $m_H=125.09\,GeV$ \cite{Aad:2015zhl}, and the top mass $m_t=173.34\,GeV$ \cite{ATLAS:2014wva}. To see the effect from varying $m_t$, we also compute $B/B_0$ for $m_t=173.34\pm 1.52\,GeV$ corresponding to the $2\sigma$ experimental uncertainty region \cite{ATLAS:2014wva}.\footnote{Within this region, $\lambda$ changes sign at some scale $h_*\ll M_P$, and the tunneling is possible.} The results for two normalization points $\hat{\mu}=\hat{\mu}_{I,II}$, with $\hat{\mu}_{I,II}$ given in (\ref{NormalizationPoints}), are presented in Fig.\ref{Plot:Bs}.

We observe that the difference between the results obtained within different normalization prescriptions is small. The behavior of $B$ as a function of the non--minimal couplings $\xi_h$ and $\xi_\chi$ confirms the predictions (\ref{ChangeOfBI}) based on the qualitative analysis of the effective potential for the bounce solution. We also see that necessarily $B>B_0$. This is to be expected, since the bounce interpolates between the (approximately) flat space and the AdS space, and the gravitational effects are known to make the transition from Minkowski geometry to AdS geometry less probable compared to the flat space limit \cite{Coleman:1980aw}.

\subsection{EW vacuum stability in the Higgs--inflation scenario}

\begin{figure}[h!]
\begin{minipage}[h]{0.44\linewidth}
\center{\includegraphics[width=1\linewidth]{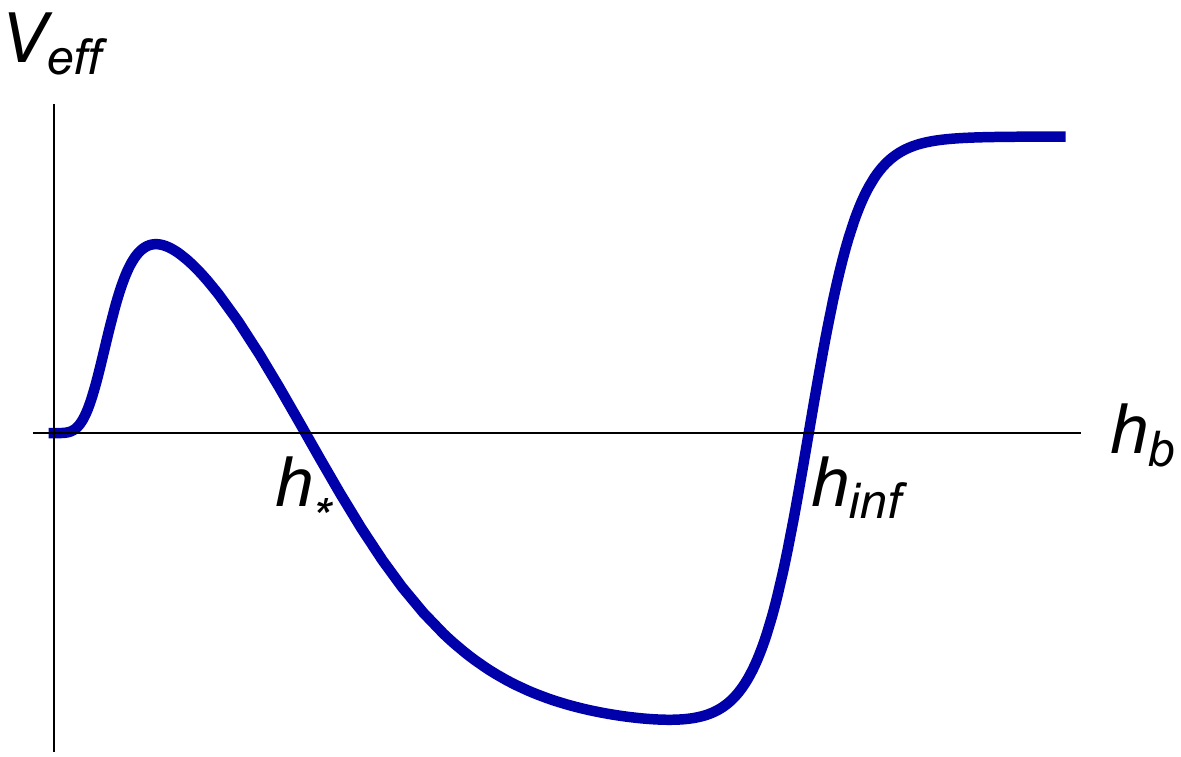}} \\(a)
\end{minipage}
\hfill
\begin{minipage}[h]{0.44\linewidth}
\center{\includegraphics[width=1\linewidth]{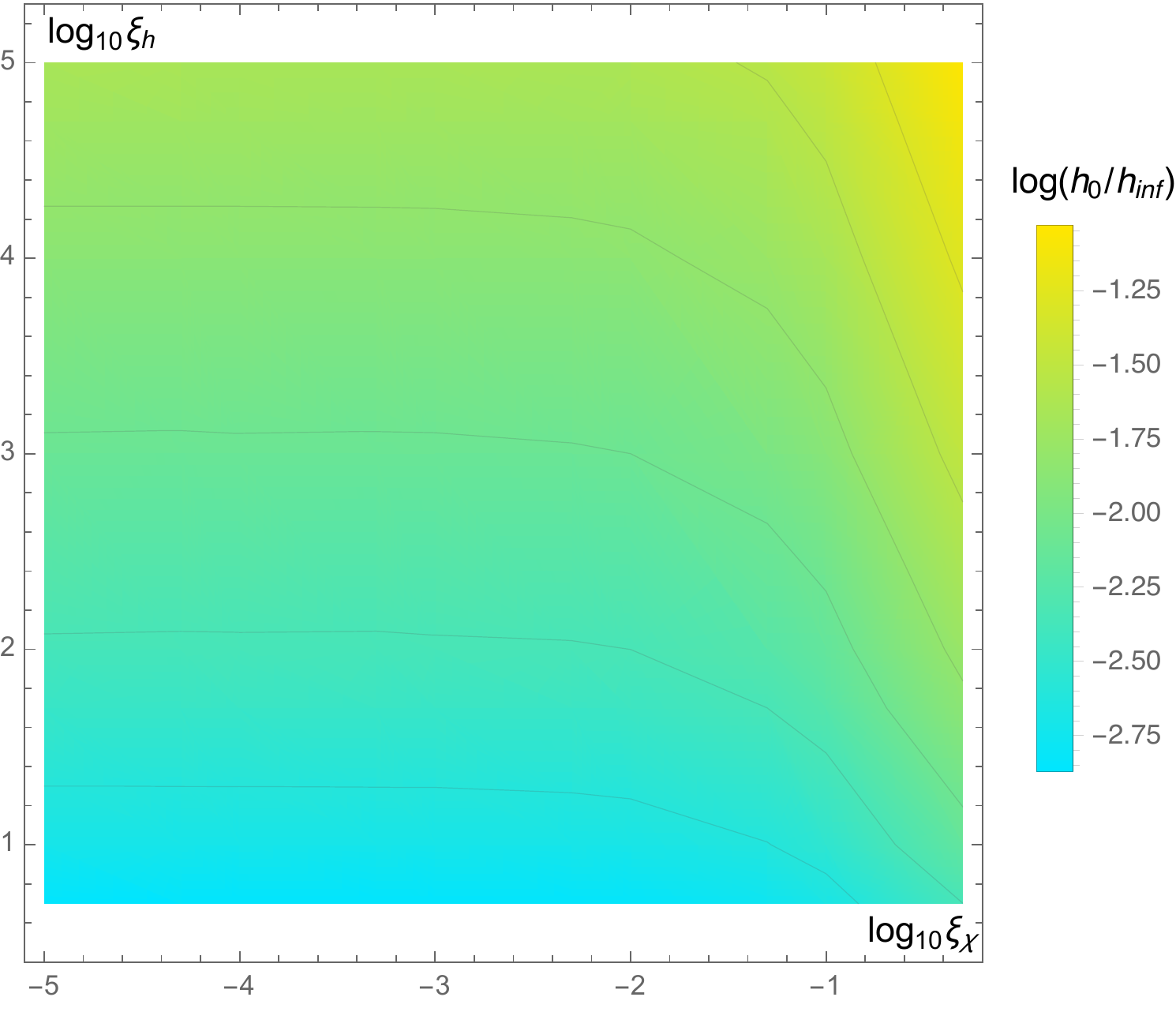}} \\(b)
\end{minipage}
\caption{(a) Schematic form of the effective potential in the Higgs--inflation scenario. (b) Magnitude of the bounce relative to the scale of the Higgs inflation. Here we take $m_t=173.34\,GeV$ and $\hat{\mu}=\hat{\mu}_I$.}
\label{Plot:Misc}
\end{figure}

Before we have discussed how the quantum corrections affect the shape of the effective potential for the bounce solution at the relevant energy scales. Let us now discuss the possibility that these corrections change the potential in the way that makes the possible metastability of the EW vacuum compatible with the Higgs--infation scenario \cite{Bezrukov:2014ipa}. Renormalization effects due to non--minimal coupling of the Higgs field can bring the Higgs self--coupling to positive values at inflationary scales, as shown schematically in Fig.\ref{Plot:Misc}(a). A typical energy at which these effects take over is of the order $h_{\textit{inf}}\sim M_P/\xi_h$. As long as this scale exceeds the magnitude of the bounce $h_0$, the corrections do not affect the decay rate. On the other hand, if $h_0\gtrsim h_{\textit{inf}}$, we expect the bounce to change significantly, yielding the further suppression of the tunneling probability. Somewhat surprisingly, numerical calculations show that for the values of $\xi_h$ and $\xi_\chi$ that we consider here, $h_0$ never approaches $h_{\textit{inf}}$. We illustrate this point in Fig.\ref{Plot:Misc}(b), where we choose, as an example, $m_t=173.34\,GeV$ and $\hat{\mu}=\hat{\mu}_I$. We conclude that the inflationary physics produces no effect on the stability of the EW vacuum in the current low--temperature background. 

\section{Discussion}

Let us first make a general comment about the bounce solution in the Higgs--Dilaton model. We would like to emphasize the fact that, according to the inequality (\ref{BounceBound}), the theory restricts the maximal scale, that the tunneling solution can hit, by the value $h_{\textit{max}}=M_P\sqrt{\frac{1+6\xi_\chi}{\xi_\chi(1+6\xi_h)}}$. This threshold is well below the Planck scale, provided that
\be\label{Ins}
\xi_\chi\xi_h\gg 1\,.
\ee 
Moreover, as is seen from the numerical findings, for the values of parameters of the theory that we discuss here, including the range of them acceptable for phenomenology, the magnitude of the bounce satisfies the much stronger condition, $h_0\ll M_P/\xi_h$. Note that the bound $h_{\textit{max}}$ is determined solely by $\xi_h$ and $\xi_\chi$ and is independent of any other couplings of the theory. Being the effective theory, the Higgs--Dilaton model possesses a UV--cutoff scale, which in our case is given by the effective Planck mass \cite{HD-2}. When approaching that scale, the theory is required to be supplemented by a sequence of higher--order operators, suppressed by the value of the cutoff. If the inequality (\ref{Ins}) holds, the introduction of these operators has no effect on the decay rate of the EW vacuum as long as they do not spoil the condition $\rho_b'=0$. This observation reveals the difference between the tunneling processes in the Higgs--Dilaton model and in the SM. Indeed, it is known that the SM--bounce can probe the sub--Planckian energies, and that the Planck--suppressed operators, added to the theory, can change drastically the predictions for the EW vacuum decay rate \cite{Branchina:2014rva}. We conclude that, compared to the case of SM, the tunneling probability in the Higgs--Dilaton model at the range of parameters specified by (\ref{Ins}) is less sensitive to new physics interfering at the Planck scales.

\begin{figure}[h]
\begin{center}
\begin{minipage}[h]{0.44\linewidth}
\center{\includegraphics[width=1\linewidth]{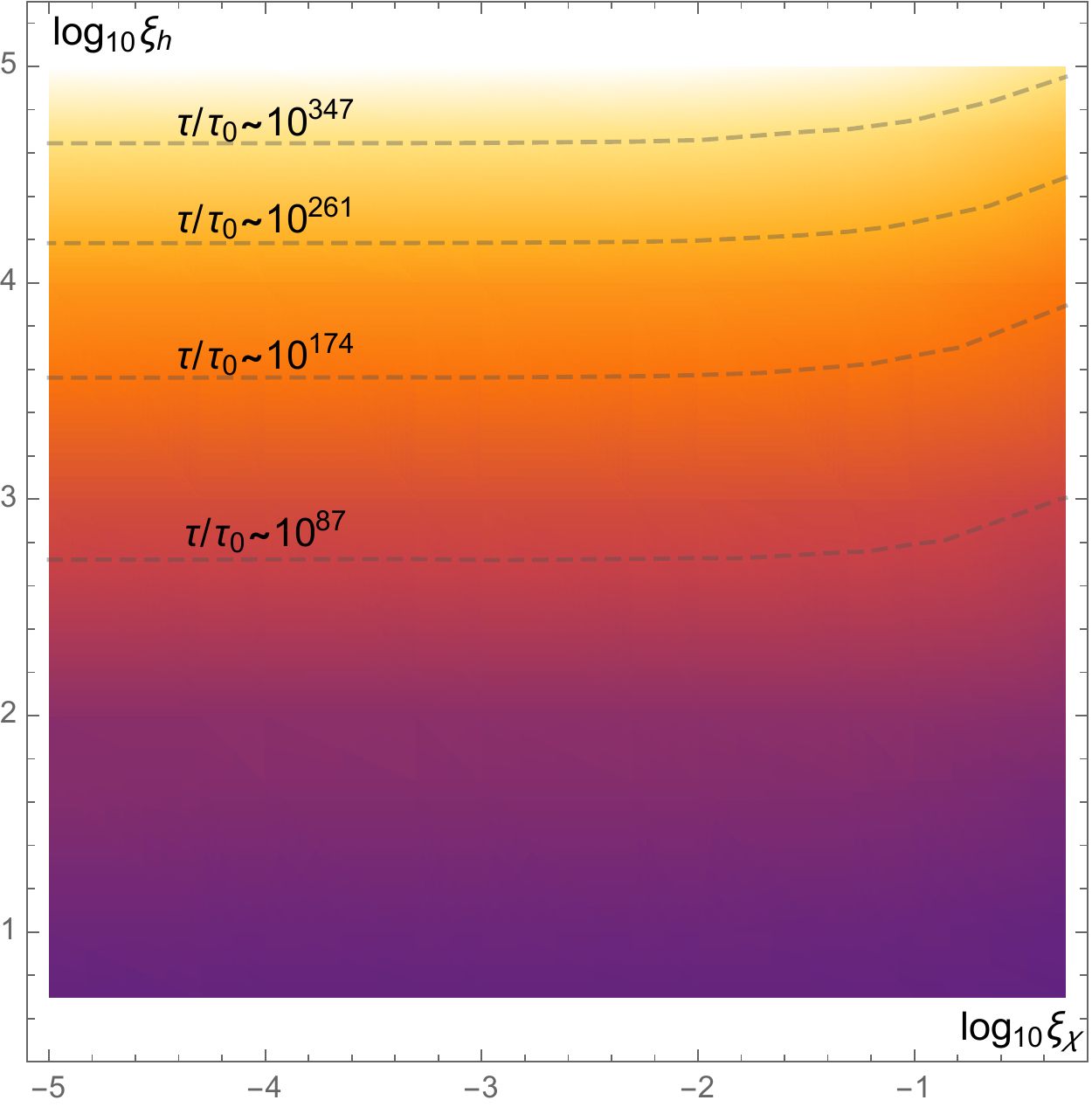}}
\end{minipage}
\caption{The ratio of the life--time of the Universe in the Higgs--Dilaton model ($\tau$) to that in the SM ($\tau_0$). Here we choose $m_t=173.34\,GeV$ and $\hat{\mu}=\hat{\mu}_I$.}
\label{Plot:LifeTime}
\end{center}
\end{figure}

Let us now discuss the life--time of the Universe in the Higgs--Dilaton model. Whenever the EW vacuum is not absolutely stable, there remains possibility for a transition towards another minimum of the Higgs potential. We would like to make sure that the calculations we have performed for the exponential factor $B$ guarantee the expected life--time to exceed the present age of the Universe by many orders of magnitude. To this end, one needs to estimate the prefactor $A$ introduced in (\ref{DecayRate}). In flat space--time, the good estimation for $A$ is \cite{Prefactor-flat-SM}
\be \label{Prefactor}
A\sim R^{-4}\,,
\ee
where $R$ is the full--width--half--maximum of the bounce. We assume that the formula (\ref{Prefactor}) remains valid after gravity and non--minimal couplings are taken into account. Then, the life--time is given by \cite{Branchina:2014rva}
\be 
\tau=\dfrac{R^4}{T_U^3}e^{B}\,,
\ee
where $T_U$ is the age of the Universe. For example, taking $m_H=125.09\,GeV$ and $m_t=173.34\,GeV$, we have in the SM case \cite{Branchina:2015nda}
\be 
\tau_0\sim 10^{600}T_U\,.
\ee
One can estimate the additional suppression of the decay rate in the Higgs--Dilaton model by computing the ratio $\tau/\tau_0$ for different couplings $\xi_h$ and $\xi_\chi$. As an example, Fig.\ref{Plot:LifeTime} shows the ratio when choosing $m_t=173.34\,GeV$ and $\hat{\mu}=\hat{\mu}_I$. In particular, we observe that for the values of $\xi_h$ acceptable for inflation, the life--time of the EW vacuum is enhanced by at least $130$ orders of magnitude compared to the SM case. Thus, the EW vacuum of the Higgs--Dilaton model is much safer than that of the SM.

\section{Conclusion}

In this paper we have investigated the EW vacuum decay in the Higgs--Dilaton model. We have addressed the question of vacuum stability for the wide range of parameters of the theory. The stability of the EW vacuum against the transition towards another minimum of the Higgs potential is one of the necessary ingredients that make the theory phenomenologically viable. Our analysis showed that the transition probability is suppressed significantly compared to the SM case, yielding the further stabilization of the EW vacuum. We also pointed out that possible corrections to the Higgs potential, coming from inflationary physics, do not change the life--time of the EW vacuum. Furthermore, the decay rate of the EW vacuum in the Higgs--Dilaton model is less sensitive to the higher--order Planck--suppressed operators, than the decay rate in the SM, provided that $\xi_h\xi_\chi\gg 1$.  

\paragraph{Acknowledgments} The author thanks Mikhail Shaposhnikov and Sergey Sibiryakov for useful discussions. The work was supported by the Swiss National Foundation grant No.200020\_162927/1.

\bibliography{refHDbounce}

\providecommand{\href}[2]{#2}\begingroup\raggedright\begin{thebibliography}{10}

\bibitem{Bezrukov:2012sa}
F.~Bezrukov, M.~Y. Kalmykov, B.~A. Kniehl, and M.~Shaposhnikov, ``{Higgs Boson
  Mass and New Physics},''
  \href{http://dx.doi.org/10.1007/JHEP10(2012)140}{{\em JHEP} {\bfseries 10}
  (2012) 140},
\href{http://arxiv.org/abs/1205.2893}{{\ttfamily arXiv:1205.2893 [hep-ph]}}.

\bibitem{Degrassi:2012ry}
G.~Degrassi, S.~Di~Vita, J.~Elias-Miro, J.~R. Espinosa, G.~F. Giudice,
  G.~Isidori, and A.~Strumia, ``{Higgs mass and vacuum stability in the
  Standard Model at NNLO},''
  \href{http://dx.doi.org/10.1007/JHEP08(2012)098}{{\em JHEP} {\bfseries 08}
  (2012) 098},
\href{http://arxiv.org/abs/1205.6497}{{\ttfamily arXiv:1205.6497 [hep-ph]}}.

\bibitem{Buttazzo:2013uya}
D.~Buttazzo, G.~Degrassi, P.~P. Giardino, G.~F. Giudice, F.~Sala, A.~Salvio,
  and A.~Strumia, ``{Investigating the near-criticality of the Higgs boson},''
  \href{http://dx.doi.org/10.1007/JHEP12(2013)089}{{\em JHEP} {\bfseries 12}
  (2013) 089},
\href{http://arxiv.org/abs/1307.3536}{{\ttfamily arXiv:1307.3536 [hep-ph]}}.

\bibitem{Why-should-we-care}
F.~Bezrukov and M.~Shaposhnikov, ``{Why should we care about the top quark
  Yukawa coupling?},'' \href{http://dx.doi.org/10.1134/S1063776115030152}{{\em
  J. Exp. Theor. Phys.} {\bfseries 120} (2015) 335--343},
  \href{http://arxiv.org/abs/1411.1923}{{\ttfamily arXiv:1411.1923 [hep-ph]}}.
[Zh. Eksp. Teor. Fiz.147,389(2015)].

\bibitem{Spannagel:2016cqt}
S.~Spannagel, ``{Top quark mass measurements with the CMS experiment at the
  LHC},'' {\em PoS} {\bfseries DIS2016} (2016) 150,
\href{http://arxiv.org/abs/1607.04972}{{\ttfamily arXiv:1607.04972 [hep-ex]}}.

\bibitem{YauWong:2016idk}
{\bfseries ATLAS} Collaboration, K.~Yau~Wong, ``{New results on top-quark mass,
  including new methods, in ATLAS},'' in {\em {9th International Workshop on
  Top Quark Physics (TOP 2016) Olomouc, Czech Republic, September 19-23,
  2016}}.
\newblock 2016.
\newblock \href{http://arxiv.org/abs/1610.06707}{{\ttfamily arXiv:1610.06707
  [hep-ex]}}.
\newblock
\url{http://inspirehep.net/record/1493844/files/arXiv:1610.06707.pdf}.
\newblock

\bibitem{ATLAS:2014wva}
{\bfseries ATLAS, CDF, CMS, D0} Collaboration, ``{First combination of Tevatron
  and LHC measurements of the top-quark mass},''
\href{http://arxiv.org/abs/1403.4427}{{\ttfamily arXiv:1403.4427 [hep-ex]}}.

\bibitem{Aad:2015zhl}
{\bfseries ATLAS, CMS} Collaboration, G.~Aad {\em et~al.}, ``{Combined
  Measurement of the Higgs Boson Mass in $pp$ Collisions at $\sqrt{s}=7$ and 8
  TeV with the ATLAS and CMS Experiments},''
  \href{http://dx.doi.org/10.1103/PhysRevLett.114.191803}{{\em Phys. Rev.
  Lett.} {\bfseries 114} (2015) 191803},
\href{http://arxiv.org/abs/1503.07589}{{\ttfamily arXiv:1503.07589 [hep-ex]}}.

\bibitem{GravCorr-SM-1}
G.~Isidori, V.~S. Rychkov, A.~Strumia, and N.~Tetradis, ``{Gravitational
  corrections to standard model vacuum decay},''
  \href{http://dx.doi.org/10.1103/PhysRevD.77.025034}{{\em Phys. Rev.}
  {\bfseries D77} (2008) 025034},
\href{http://arxiv.org/abs/0712.0242}{{\ttfamily arXiv:0712.0242 [hep-ph]}}.

\bibitem{GravCorr-SM-2}
V.~Branchina, E.~Messina, and D.~Zappala, ``{Impact of Gravity on Vacuum
  Stability},''
\href{http://arxiv.org/abs/1601.06963}{{\ttfamily arXiv:1601.06963 [hep-ph]}}.

\bibitem{Rajantie:2016hkj}
A.~Rajantie and S.~Stopyra, ``{Standard Model vacuum decay with gravity},''
\href{http://arxiv.org/abs/1606.00849}{{\ttfamily arXiv:1606.00849 [hep-th]}}.

\bibitem{Salvio:2016mvj}
A.~Salvio, A.~Strumia, N.~Tetradis, and A.~Urbano, ``{On gravitational and
  thermal corrections to vacuum decay},''
  \href{http://dx.doi.org/10.1007/JHEP09(2016)054}{{\em JHEP} {\bfseries 09}
  (2016) 054},
\href{http://arxiv.org/abs/1608.02555}{{\ttfamily arXiv:1608.02555 [hep-ph]}}.

\bibitem{Cosmology-1}
J.~R. Espinosa, G.~F. Giudice, and A.~Riotto, ``{Cosmological implications of
  the Higgs mass measurement},''
  \href{http://dx.doi.org/10.1088/1475-7516/2008/05/002}{{\em JCAP} {\bfseries
  0805} (2008) 002},
\href{http://arxiv.org/abs/0710.2484}{{\ttfamily arXiv:0710.2484 [hep-ph]}}.

\bibitem{Cosmology-2}
O.~Lebedev and A.~Westphal, ``{Metastable Electroweak Vacuum: Implications for
  Inflation},'' \href{http://dx.doi.org/10.1016/j.physletb.2012.12.069}{{\em
  Phys. Lett.} {\bfseries B719} (2013) 415--418},
\href{http://arxiv.org/abs/1210.6987}{{\ttfamily arXiv:1210.6987 [hep-ph]}}.

\bibitem{Cosmology-3}
M.~Herranen, T.~Markkanen, S.~Nurmi, and A.~Rajantie, ``{Spacetime curvature
  and Higgs stability after inflation},''
  \href{http://dx.doi.org/10.1103/PhysRevLett.115.241301}{{\em Phys. Rev.
  Lett.} {\bfseries 115} (2015) 241301},
\href{http://arxiv.org/abs/1506.04065}{{\ttfamily arXiv:1506.04065 [hep-ph]}}.

\bibitem{Cosmology-4}
A.~Shkerin and S.~Sibiryakov, ``{On stability of electroweak vacuum during
  inflation},'' \href{http://dx.doi.org/10.1016/j.physletb.2015.05.012}{{\em
  Phys. Lett.} {\bfseries B746} (2015) 257--260},
\href{http://arxiv.org/abs/1503.02586}{{\ttfamily arXiv:1503.02586 [hep-ph]}}.

\bibitem{BH-1}
P.~Burda, R.~Gregory, and I.~Moss, ``{Gravity and the stability of the Higgs
  vacuum},'' \href{http://dx.doi.org/10.1103/PhysRevLett.115.071303}{{\em Phys.
  Rev. Lett.} {\bfseries 115} (2015) 071303},
\href{http://arxiv.org/abs/1501.04937}{{\ttfamily arXiv:1501.04937 [hep-th]}}.

\bibitem{BH-2}
P.~Burda, R.~Gregory, and I.~Moss, ``{The fate of the Higgs vacuum},''
  \href{http://dx.doi.org/10.1007/JHEP06(2016)025}{{\em JHEP} {\bfseries 06}
  (2016) 025},
\href{http://arxiv.org/abs/1601.02152}{{\ttfamily arXiv:1601.02152 [hep-th]}}.

\bibitem{HD-1}
J.~Garcia-Bellido, J.~Rubio, M.~Shaposhnikov, and D.~Zenhausern,
  ``{Higgs-Dilaton Cosmology: From the Early to the Late Universe},''
  \href{http://dx.doi.org/10.1103/PhysRevD.84.123504}{{\em Phys. Rev.}
  {\bfseries D84} (2011) 123504},
\href{http://arxiv.org/abs/1107.2163}{{\ttfamily arXiv:1107.2163 [hep-ph]}}.

\bibitem{HD-2}
F.~Bezrukov, G.~K. Karananas, J.~Rubio, and M.~Shaposhnikov, ``{Higgs-Dilaton
  Cosmology: an effective field theory approach},''
  \href{http://dx.doi.org/10.1103/PhysRevD.87.096001}{{\em Phys. Rev.}
  {\bfseries D87} no.~9, (2013) 096001},
\href{http://arxiv.org/abs/1212.4148}{{\ttfamily arXiv:1212.4148 [hep-ph]}}.

\bibitem{Coleman:1977py}
S.~R. Coleman, ``{The Fate of the False Vacuum. 1. Semiclassical Theory},''
  \href{http://dx.doi.org/10.1103/PhysRevD.15.2929,
  10.1103/PhysRevD.16.1248}{{\em Phys. Rev.} {\bfseries D15} (1977)
  2929--2936}.
[Erratum: Phys. Rev.D16,1248(1977)].

\bibitem{Prefactor-flat-SM}
G.~Isidori, G.~Ridolfi, and A.~Strumia, ``{On the metastability of the standard
  model vacuum},'' \href{http://dx.doi.org/10.1016/S0550-3213(01)00302-9}{{\em
  Nucl. Phys.} {\bfseries B609} (2001) 387--409},
\href{http://arxiv.org/abs/hep-ph/0104016}{{\ttfamily arXiv:hep-ph/0104016
  [hep-ph]}}.

\bibitem{Gibbons:1976ue}
G.~W. Gibbons and S.~W. Hawking, ``{Action Integrals and Partition Functions in
  Quantum Gravity},''
\href{http://dx.doi.org/10.1103/PhysRevD.15.2752}{{\em Phys. Rev.} {\bfseries
  D15} (1977) 2752--2756}.

\bibitem{Masoumi:2016pqb}
A.~Masoumi, S.~Paban, and E.~J. Weinberg, ``{Tunneling from a Minkowski vacuum
  to an AdS vacuum: A new thin-wall regime},''
  \href{http://dx.doi.org/10.1103/PhysRevD.94.025023}{{\em Phys. Rev.}
  {\bfseries D94} no.~2, (2016) 025023},
\href{http://arxiv.org/abs/1603.07679}{{\ttfamily arXiv:1603.07679 [hep-th]}}.

\bibitem{Fujii:2003pa}
Y.~Fujii and K.~Maeda, {\em {The scalar-tensor theory of gravitation}}.
\newblock Cambridge University Press, 2007.
\newblock
\url{http://www.cambridge.org/uk/catalogue/catalogue.asp?isbn=0521811597}.
\newblock

\bibitem{O4-flat}
S.~R. Coleman, V.~Glaser, and A.~Martin, ``{Action Minima Among Solutions to a
  Class of Euclidean Scalar Field Equations},''
\href{http://dx.doi.org/10.1007/BF01609421}{{\em Commun. Math. Phys.}
  {\bfseries 58} (1978) 211}.

\bibitem{Blum:2016ipp}
K.~Blum, M.~Honda, R.~Sato, M.~Takimoto, and K.~Tobioka, ``{O($N$) Invariance
  of the Multi-Field Bounce},''
\href{http://arxiv.org/abs/1611.04570}{{\ttfamily arXiv:1611.04570 [hep-th]}}.

\bibitem{SI-renorm}
M.~Shaposhnikov and D.~Zenhausern, ``{Quantum scale invariance, cosmological
  constant and hierarchy problem},''
  \href{http://dx.doi.org/10.1016/j.physletb.2008.11.041}{{\em Phys. Lett.}
  {\bfseries B671} (2009) 162--166},
\href{http://arxiv.org/abs/0809.3406}{{\ttfamily arXiv:0809.3406 [hep-th]}}.

\bibitem{SI-renorm-2}
F.~Englert, C.~Truffin, and R.~Gastmans, ``{Conformal Invariance in Quantum
  Gravity},''
\href{http://dx.doi.org/10.1016/0550-3213(76)90406-5}{{\em Nucl. Phys.}
  {\bfseries B117} (1976) 407--432}.

\bibitem{RG-code-based}
K.~G. Chetyrkin and M.~F. Zoller, ``{Three-loop $\beta$-functions for
  top-Yukawa and the Higgs self-interaction in the Standard Model},''
  \href{http://dx.doi.org/10.1007/JHEP06(2012)033}{{\em JHEP} {\bfseries 06}
  (2012) 033},
\href{http://arxiv.org/abs/1205.2892}{{\ttfamily arXiv:1205.2892 [hep-ph]}}.

\bibitem{Coleman:1980aw}
S.~R. Coleman and F.~De~Luccia, ``{Gravitational Effects on and of Vacuum
  Decay},''
\href{http://dx.doi.org/10.1103/PhysRevD.21.3305}{{\em Phys. Rev.} {\bfseries
  D21} (1980) 3305}.

\bibitem{Bezrukov:2014ipa}
F.~Bezrukov, J.~Rubio, and M.~Shaposhnikov, ``{Living beyond the edge: Higgs
  inflation and vacuum metastability},''
  \href{http://dx.doi.org/10.1103/PhysRevD.92.083512}{{\em Phys. Rev.}
  {\bfseries D92} no.~8, (2015) 083512},
\href{http://arxiv.org/abs/1412.3811}{{\ttfamily arXiv:1412.3811 [hep-ph]}}.

\bibitem{Branchina:2014rva}
V.~Branchina, E.~Messina, and M.~Sher, ``{Lifetime of the electroweak vacuum
  and sensitivity to Planck scale physics},''
  \href{http://dx.doi.org/10.1103/PhysRevD.91.013003}{{\em Phys. Rev.}
  {\bfseries D91} (2015) 013003},
\href{http://arxiv.org/abs/1408.5302}{{\ttfamily arXiv:1408.5302 [hep-ph]}}.

\bibitem{Branchina:2015nda}
V.~Branchina and E.~Messina, ``{Stability and UV completion of the Standard
  Model},''
\href{http://arxiv.org/abs/1507.08812}{{\ttfamily arXiv:1507.08812 [hep-ph]}}.

\end{thebibliography}\endgroup

\end{document}